\documentclass{aa}
\newcommand{\kms}{{ km s}$^{-1}$}
\newcommand{\sci}{{\,{\sc i}}}
\newcommand{\scii}{{\,{\sc ii}}}
\newcommand{\vsini}{{$v\sin i$}}
\newcommand{\amd}{{1991, in: The Sun and Cool Stars, IAU Coll. 130,
Tuominen I., Moss D., R\"udiger G. (eds.), Springer-Verlag, Heidelberg}}
\begin{document}
\title{Surface imaging of \object{HD~199178}~(\object{V1794 Cygni}) 
\thanks{Based on observations made with the 2 m telescope of the National 
Astronomical Observatory, Rozhen, Bulgaria, and the Nordic Optical Telescope, 
operated on the island of La Palma jointly by Denmark, Finland, Iceland, 
Norway, and Sweden, in the Spanish Observatorio del Roque de los Muchachos of 
the Instituto de Astrofisica de Canarias.}}
\titlerunning{Surface imaging of \object{HD~199178}}
\author{T. Hackman \inst{1} 
\and L. Jetsu \inst{1}
\and I. Tuominen \inst{2}}
\offprints{Thomas.Hackman@Helsinki.Fi}
\institute{Observatory, P.O. Box 14,  FIN-00014 University of Helsinki, 
           Finland
\and
           Astronomy Division, P.O. Box 3000, FIN-90014 University of
           Oulu, Finland}
\date{Received; accepted }

\abstract{ 
We present surface temperature maps for the FK Comae-type star
\object{HD~199178} (V1794 Cygni) calculated from high resolution spectra 
obtained in 1994 and 1995. The spot pattern evolves, but all maps
reveal a large cool spot remaining nearly at the same high latitude. 
The main spot is 1200 -- 1600 K cooler than the mean surface
temperature. The observed slightly flat bottomed absorption lines
would usually be interpreted as evidence for a large cool polar
spot. We argue that antisolar surface differential rotation offers a
better explanation for the box like shape of the line profiles. However, we do 
not find conclusive evidence for antisolar differential rotation and note
that there are still other possible explanations for the slightly flat 
bottomed line profiles.
}

\maketitle

\section{Introduction}

Doppler or surface imaging is, at present, the most powerful tool to
map spots on rapidly rotating late-type stars. Even so, the Doppler
imaging maps should always be studied critically. The method suffers
from difficulties in modeling spectral lines in late-type stars, as
well as from uncertainties in stellar and spectral parameters. These
problems have been amply illustrated by conflicting results. For
instance, the temperature maps obtained with some Doppler imaging
methods most often show very large cool spots covering the poles 
(cf. Vogt \cite{vogt1}; Strassmeier \cite{hd106225};
Hatzes \& Vogt \cite{hv}), while the results with the methods
developed by Piskunov (\cite{pisk}) and Berdyugina (\cite{occam})
rarely show polar spots (cf. Piskunov et al. \cite{si1}; 
Strassmeier et al. \cite{eieri}; Korhonen et al. \cite{heidi2}). Since 
partly the same stars, and once even the same observations
(Strassmeier et al. \cite{eieri}), have been modeled, it is evident
that the results depend either on the applied methods or
the selected stellar and spectral parameters. 

There are several studies on how line and stellar parameters influence
Doppler imaging (cf. Unruh \& Collier Cameron \cite{unruh}). 
But the influence of one important phenomenon, namely surface 
differential rotation, has usually been neglected. Many active late-type stars
show clear variations in their photometric rotation periods (Strassmeier et 
al. \cite{klasu1}; Cutispoto \cite{cuti}). Variable periods have
also been discovered in the Ca\scii~H\&K emission (Donahue et
al. \cite{donahue}). These periodicity changes are usually explained as
a signature of surface differential rotation combined with latitudinal 
activity migration. Apart from distorting the spot pattern, differential
rotation would alter the shape of the line profile in the same way
as a large cool or hot polar spot. This emphasizes the necessity of
implementing surface differential rotation into the Doppler imaging
techniques.

The generally accepted empirical result is that differential
rotation is weaker in more rapidly rotating stars (Hall
\cite{hall}). Hence the effects of differential rotation have usually been
bypassed in Doppler imaging. Although antisolar differential
rotation has previously been noted to alter the line profiles in a similar 
way as a large cool polar spot, Hatzes et al. (\cite{polspot}) concluded
that this would require an unreasonably strong differential rotation.

\object{HD~199178} (V1794 Cygni) is a rapidly rotating single G5 subgiant 
(Herbig \cite{herbig}), which was classified as a FK Comae-type star by 
Bopp \& Rucinski (\cite{bopp1}). Jetsu et al. (\cite{jetsu1}) derived a 
photometric rotation period of $P_{\rm phot} = 3\fd337484$ $\pm 
0\fd000043$. Furthermore, they reported a 9.1 year period in the mean 
brightness level and indications of a shorter cycle in the light curve 
amplitude. In a recent analysis of photometry between 1975 and 1995, 
Jetsu et al. (\cite{time}) showed that the photometric period varied and 
interpreted this as a signature of differential rotation.

We present new temperature maps of \object{HD~199178} for the years
1994 and 1995. The maps were calculated with the surface imaging technique 
developed by Piskunov (\cite{pisk}). Previous Doppler images of this star 
display a large cool polar spot and some smaller high and low latitude 
features (Vogt \cite{vogt1}; Strassmeier et al. \cite{klasu2}). Unlike these 
earlier studies of \object{HD~199178}, we take into account the possible 
effect of differential rotation on the rotational broadening of photospheric 
absorption lines. We also demonstrate how different values for the 
differential rotation coefficient alter the retrieved surface imaging map.

\section{Stellar parameters of HD~199178}

\object{HD~199178} has an apparent visual magnitude of $V \approx
7\fm2$. The Hipparcos-satellite parallax measurement of $\pi
\approx 10.68\!\pm 0.73$  mas gives a distance of 94 pc (Perryman 
\cite{hipp}). The long-term mean colour indices
$B\!-V\!\approx 0.79$ and $V\!-\!R \approx 0.68 $ determined by Jetsu et
al. (\cite{jetsu1}) suggested an average ``background'' stellar surface 
temperature of $T_{\rm eff} \approx$ 5375 K. The above results agree with the 
G5{\sc iii}-{\sc iv} spectral classification by Herbig (\cite{herbig}).

The new Hipparcos distance measurement allows us to estimate the
absolute magnitude and radius of the star. We use $7\fm09$ as
the apparent visual magnitude of the unspotted star (Strassmeier et al. 
\cite{klasu2}). \object{HD~199178} lies in the direction of the
North America and Pelican Nebulae complex, but in the foreground.
Assuming an interstellar absorption of $0\fm2$ (Strai{\u z}ys et 
al. \cite{ext}: Fig. 2) yields an absolute
visual magnitude of $2\fm02$. A bolometric correction of $BC=-0.184$
(Strassmeier et al. \cite{klasu2}) gives the bolometric magnitude
$M_{\rm bol} \approx 1\fm84$. Finally, if the unspotted temperature of
$T_{\rm eff}=5400$ K is combined to the solar $T_\odot = 5800$ K and 
$M_{\rm bol_\odot}= 4\fm64$, the Stefan-Boltzmann relation
gives $R \approx 4.2 R_\odot$.

The Barnes-Evans relation is based on the same principle as
the above, except that the value of the bolometric correction is not
required. We use the formulation by Lacy (\cite{lacy}):

\begin{equation}
\lg (R/R_{\odot}) = 7.4724 - 0.2 V_0 - 2 F_v + \lg d,
\end{equation}

\noindent
where $F_v = 3.977 - 0.429 (V-R)_0$ and $[d]=$ pc. The combination
$V_0 = 7\fm0$, $V-R = 0.68$ and $E_{V-R}=0.05$ gives $F_v \approx
3.707$ and $R \approx (4.3 \pm 0.7) R_\odot$. (Here $V_0$ is the mean apparent 
visual magnitude without interstellar absorption.)

The rotation period $P \approx 3\fd3$ and projected rotational velocity 
$v\sin i$ $\approx$ 70 \kms~ give a lower limit $R \sin i \approx 4.6 
R_\odot$ for the radius. There is thus a slight discrepancy between the 
different radius determinations. Anyway, the above estimates and the lower 
limit $R \sin i$ suggest that $4.6 R_\odot \la R \la 5.0 R_\odot$. This is
equivalent to an inclination of $67^{\circ} \la i \la 90^{\circ}$.
The exceptional nature of \object{HD~199178} (rapid rotation, spots, possibly 
a coalesced binary) may mean that the equatorial radius will be 
underestimated. In our calculations we will thus allow for an inclination 
$i \geq 50^{\circ}$. 

A variable photometric rotation period was detected in the long-term 
photometry by Jetsu et al. (\cite{time}). They analyzed 114
sets of photometric data obtained between 1975 and 1995. Each set contained
at least 8 nights of observations and had a length of about 30 nights (Jetsu et
al. \cite{photo}). The regular yearly change rate of the photometric period 
was 3.3 \%. In addition, irregular seasonal changes of 7.5 \% were found. 
The varying photometric period was explained by surface differential rotation 
quantified by $ \vert \alpha \vert \ga 0.075 $, where 

\begin{equation}
\alpha = (\Omega_{\rm equator}- \Omega_{\rm pole}) / \Omega_{\rm
 equator}.
\label{eq:dr}
\end{equation}

\noindent
Even this lower limit for the differential rotation is considerably larger
than that expected from the results by Hall (\cite{hall}).
If the latitudinal range of the spot activity does not extend all the way from
the equator to the pole, then $\vert \alpha \vert$ could even reach the solar 
value of 0.2. We thus assume that the realistic differential rotation 
coefficient values are within the limits $-0.20 \le \alpha  \le 0.20$.

Differential rotation would alter the shape of the line profile. The case 
of $\alpha > 0$ induces a ``sharper'' profile, while $\alpha < 0$ causes a
more ``flat bottomed'' profile (cf. Bruning \cite{bruning}). Thus a
negative $\alpha$ modifies the line profile much in the same way as a
cool polar spot.

Since there is no unique photometric period, in principle different
periods could be used for different seasons. But this would make it
harder to compare the surface images. Therefore we chose to use the mean
period $P= 3\fd3250$ calculated for the years 1994 an 1995. Our ephemeris
for the rotational phases was

\begin{equation}
{\rm HJD_{min}}\!=\!(2449511.150 \pm 0.070)\!+\!(3.3250 \pm 0.0012) E,
\label{eq:eph}
\end{equation}

\noindent
where the ${\rm HJD_{min}}$ is the epoch of the light curve minimum
(Jetsu et al. \cite{time}).

\section{The surface imaging technique}

Spots on the surface of a rapidly rotating star will cause distortions 
(``bumps'') in the observed spectral line profiles. As the star rotates, these 
bumps will move across the absorption line profiles. The main idea of 
surface imaging is to trace these distortions and retrieve a surface map of 
the star. We use the surface imaging technique developed by Piskunov
(Piskunov et al. \cite{si1}, Eqs. 3--6 and 8;  Piskunov \cite{pisk},
Eqs. 5--8), but also include Johnson $V$-photometry in the 
inversion.

Local line profiles and $V$-band fluxes are calculated for a set
of temperatures and limb angles using stellar model atmospheres. This table 
is then used for the disk integration of a given surface temperature 
distribution. The surface imaging problem is solved by searching for a 
surface temperature distribution that minimizes the discprepancy between the 
observations and the calculated line profiles and light curve.

Since we always have a confined temperature range of model atmospheres,
the solution should be limited to temperatures between
the minimum and maximum of the models ($T_{\rm min}$ and $T_{\rm
max}$). We do this by adding a ``penalty function'' to the quantity to
be minimized. We use an integral of a double-sided Heaviside step
function, which is nearly constant between $T_{\rm min}$ and 
$T_{\rm max}$, but grows nearly linearly when the temperature is below 
$T_{\rm min}$ or exceeds $T_{\rm max}$. Thus the penalty function merely 
prevents extreme temperatures in the solution.

The crucial point of surface imaging is how the line profiles for a given 
temperature distribution are calculated. Errors in the calculated line
profiles will introduce artifacts in the resulting map and prevent the
solution from reaching the observational signal-to-noise level. Errors
in the line profiles can arise from numerous problems, e.g. wrong
line parameters, missing lines, non-LTE effects, wrong limb darkening or
inadequacies of the model atmospheres. The fact that the lines used for
Doppler imaging are strong absorption lines, whose cores are formed high in the
photosphere, puts extra demands on the line modeling.

The errors in the line profiles are not the only source of systematic
errors. For example, an erroneous $v\sin i$ will manifest itself as a 
cool or hot belt in the image (cf. Berdyugina \cite{occam}). Errors in the
microturbulence velocity will mainly change the average temperature of the
solution. Doppler images are not sensitive to errors in the macroturbulence 
velocity (cf. Unruh \& Collier Cameron \cite{unruh}). A wrong inclination 
will cause a latitudinal displacement and a slight distortion of the spot. 
Combined with surface differential rotation, the inclination will also 
influence the line profile. Decreasing the inclination will increase the
effect of differential rotation on the line profile (cf. Bruning 
\cite{bruning}).

Differential rotation will, of course, also change the spot
distribution, since different latitudes have different rotational periods. In
this study we do not take this effect into account. Firstly, 
differential rotation becomes important in this sense only if the locations 
of spots extend over a large latitude range. Secondly, we assume
that the magnetic field strength is sufficient to
resist the shearing effect of differential rotation within the spot(s).

Since it is not (yet) possible to calculate perfect line profiles, one
can always expect some artifacts in Doppler images. Fortunately, most of 
these artifacts can easily be identified, since certain shapes of artifacts 
are related to particular error sources. The same error in the overall shape 
of all calculated line profiles may manifest itself as a spot band at
a certain latitude (or a polar spot). However, if the rotational phase 
coverage is inadequate, this artifact might not be axisymmetric. Minor errors 
in the calculations (e.g. missing lines in blends) or the data reduction 
(e.g. remnants of cosmic spikes or uncertainty in the continuum 
normalization) may cause thin high contrast features, like vertical
stripes, arches or even ovals. For example, the remains of a cosmic
spike in one spectrum may cause a thin high contrast artifact at a stellar 
longitude corresponding to the wavelength of the spike and rotation phase of 
the observation. This in turn might cause additional hot and cold features to 
cancel out the effect of the artifact in the remaining observed phases. A 
missing line, on the other hand, may lead to an artifact repeated at each 
observed phase. Noise in the spectra might be interpreted as a series of small
``bumps''. Because such bumps will appear at arbitrary wavelengths, 
the resulting artifacts cannot be latitudinally confined, and will therefore 
be seen as vertical stripes in the image.

It is clear, that the retrieved longitude for the spots will usually be more 
reliable than the latitude. In some cases the longitude of the spot can be
estimated even with a single spectrum, while the determination of the 
latitude requires a good phase coverage. Moreover, the latitude is sensitive 
to errors in the inclination and the general shape of the rotationally 
broadened line profile. The influence of all error sources is greatly reduced 
by using many spectral lines and simultaneous photometry in the inversion.

\begin{table}[ht]
\caption{The spectral observations of \object{HD~199178}: $t\!=\!{\rm
HJD}\!-\!2440000$ for the middle of the exposures, rotational phases
calculated from the ephemeris of Eq.~(\ref{eq:eph}),
$S/N$ ratios and the resolution element $W_\lambda$}  

\begin{center}
\begin{tabular}{ccc|ccc}

\multicolumn{6}{l}{July 1994, Rozhen, $W_\lambda$=0.20 \AA} \\

\hline
$t$    &Phase  &S/N & $t$    &Phase  &S/N  \\
\hline
  9552.341 &  0.388 &  244 & 9559.479 &  0.535 &  212 \\
  9552.457 &  0.423 &  268 & 9560.303 &  0.783 &  171 \\
  9552.576 &  0.459 &  214 & 9560.474 &  0.834 &  320 \\
  9553.309 &  0.679 &  219 & 9561.350 &  0.098 &  162 \\
  9554.559 &  0.055 &  251 & 9562.353 &  0.399 &  286 \\
  9555.421 &  0.315 &  251 & 9562.537 &  0.455 &  206 \\
  9556.506 &  0.641 &  223 & 9563.352 &  0.700 &  236 \\
  9557.537 &  0.951 &  248 & 9564.393 &  0.013 &  316 \\
  9558.346 &  0.194 &  276 & 9565.381 &  0.310 &  263 \\
  9558.561 &  0.259 &  178 &          &        &      \\

\hline
\\

\multicolumn{6}{l}{August 1994, NOT-SOFIN/ 2nd camera,$W_\lambda$=0.10 \AA} \\

\hline
$t$    &Phase  &S/N & $t$    &Phase  &S/N  \\
\hline
 9578.544 &  0.269 &  209 &  9584.412 &  0.034 &  284 \\
 9579.608 &  0.589 &  281 &  9585.409 &  0.334 &  238 \\
 9580.521 &  0.863 &  277 &  9586.410 &  0.635 &  308 \\
 9581.503 &  0.159 &  262 &  9587.441 &  0.945 &  263 \\
 9582.418 &  0.434 &  246 &  9588.424 &  0.240 &  263 \\
 9583.412 &  0.733 &  282 &           &        &      \\

\hline
\\

\multicolumn{6}{l}{November 1994, NOT-SOFIN/ 1st camera,$W_\lambda$=0.05
\AA} \\

\hline
$t$    &Phase  &S/N & $t$    &Phase  &S/N  \\
\hline
 9669.355 &  0.580 &  150 &  9675.323 &  0.375 &  210 \\
 9670.305 &  0.866 &  190 &  9676.322 &  0.676 &  210 \\
 9671.412 &  0.199 &  180 &  9677.328 &  0.978 &  230 \\
 9672.310 &  0.469 &  220 &  9678.325 &  0.278 &  250 \\
 9673.309 &  0.770 &  210 &           &        &      \\

\hline
\\

\multicolumn{6}{l}{July 1995, NOT-SOFIN/ 2nd camera,$W_\lambda$=0.10
\AA} \\

\hline
$t$    &Phase  &S/N & $t$    &Phase  &S/N  \\
\hline
 9910.599 &  0.135 &  295 &  9917.580 &  0.235 &  301 \\
 9912.573 &  0.729 &  454 &  9918.588 &  0.538 &  321 \\
 9913.598 &  0.037 &  332 &  9919.634 &  0.852 &  295 \\
 9914.576 &  0.331 &  376 &  9920.623 &  0.150 &  319 \\
 9915.616 &  0.644 &  325 &  9921.532 &  0.423 &  362 \\
 9916.628 &  0.948 &  281 &           &        &      \\

\hline
\end{tabular}
\end{center}
\label{obs}
\end{table}

\renewcommand{\arraystretch}{0.90}
\begin{table*}
\caption{Atomic lines, lower excitation potential and adopted $gf$ values}
\begin{center}
\begin{tabular}{lcrr|lcrr|lcrr}

\hline
Line & $\lambda$ (\AA)& $\chi_{\rm low}$ (eV) & $\lg (gf)$ & 
Line & $\lambda$ (\AA)& $\chi_{\rm low}$ (eV) & $\lg (gf)$ &
Line & $\lambda$ (\AA)& $\chi_{\rm low}$ (eV) & $\lg (gf)$ \\

\hline
Si\sci~ & 6407.070 & 5.871 & -2.200 & Sm\sci~ & 6425.905 & 0.282 & -1.815 & 
Ti\sci~ & 6435.315 &2.250 &-2.875 \\ 
Zr\sci~ & 6407.216 & 0.154 & -2.700 & Si\sci~ & 6426.291 & 5.954 & -2.470 & 
V\sci~  & 6435.490 &2.684 &-1.582 \\
Fe\scii~ & 6407.251 & 3.889 & -3.699 & Sm\scii~ & 6426.628 & 1.746 & -0.559 & 
Ce\sci~ & 6436.399 &0.467 &-0.822 \\
Si\sci~ & 6407.291 & 5.871 & -1.500 & Zr\sci~ & 6426.706 & 2.746 & -0.620 & 
Fe\sci~ & 6436.407 &4.186 &-2.360 \\
Co\sci~ & 6407.491 & 4.395 & -1.471 & Si\sci~ & 6426.911 & 5.954 & -2.900 & 
Y\sci~  & 6437.169 &2.294 &-0.520 \\
Fe\sci~ & 6407.643 & 4.076 & -3.620 & Sm\scii~ & 6428.355 & 1.370 & -1.150 &
Ti\sci~ & 6437.610 &3.407 &-1.889 \\
Fe\sci~ & 6408.018 & 3.686 & -1.018 & Nd\scii~ & 6428.645 & 0.205 & -1.831 &
Eu\scii~ & 6437.640 &1.320 & 0.427 \\
S\sci~  & 6408.157 & 7.868 & -1.510 & Ca\sci~ & 6428.815 & 4.441 & -1.184 & 
Si\sci~ & 6437.703 &5.863 &-2.210 \\
Fe\sci~ & 6408.332 & 4.386 & -3.563 & Fe\sci~ & 6429.071 & 4.294 & -3.407 & 
Ni\sci~ & 6437.992 &5.389 &-2.984 \\
Sr\sci~ & 6408.459 & 2.271 & 0.500  & Pr\scii~ & 6429.629 & 1.615 & -0.386 &
V\sci~  & 6438.088 &2.684 &-2.070 \\
Gd\sci~ & 6408.545 & 0.213 & -2.052 & Nd\sci~ & 6429.840 & 1.337 & -0.530 &
Cr\sci~ & 6438.586 &3.890 &-2.571 \\
Si\sci~ & 6408.671 & 5.984 & -3.090 & Ni\sci~ & 6429.859 & 4.167 & -2.104 &
Fe\sci~ & 6438.755 &4.435 &-2.329 \\
V\sci~  & 6408.814 & 2.616 & -1.992 & Co\sci~ & 6429.906 & 2.137 & -2.460 &
Ca\sci~ & 6439.075 &2.526 & 0.450 \\
Ti\sci~ & 6409.390 & 3.570 & -1.631 & Ca\sci~ & 6430.127 & 3.910 & -1.907 &
Ti\sci~ & 6439.326 &3.590 &-1.318 \\
V\sci~  & 6410.416 & 2.581 & -1.889 & Co\sci~ & 6430.290 & 4.049 & -1.028 &
Fe\sci~ & 6439.554 &4.473 &-4.304 \\ 
La\sci~ & 6410.975 & 0.373 & -0.950 & Nb\sci~ & 6430.443 & 0.740 & -1.230 &
Ti\sci~ & 6439.705 &0.813 &-6.458 \\
Sc\sci~ & 6410.984 & 4.967 & -2.870 & V\sci~  & 6430.472 & 1.955 & -0.600 & 
Ce\sci~ & 6439.964 &0.294 &-0.752 \\
Fe\sci~ & 6411.107 & 4.733 & -1.905 & Ca\sci~ & 6430.793 & 3.910 & -2.129 & 
Si\sci~ & 6440.566 &5.616 &-2.480 \\
V\sci~  & 6411.276 & 1.950 & -2.059 & Fe\sci~ & 6430.846 & 2.176 & -2.106 &
N\sci~  & 6440.938 &11.764 &-1.140 \\
Cr\sci~ & 6411.580 & 3.892 & -2.478 & Sm\scii~ & 6431.006 & 1.359 & -1.158 &
Ti\sci~ & 6440.947 &3.590 &-1.810 \\
Fe\sci~ & 6411.649 & 3.654 & -0.645 & Ca\sci~ & 6431.099 & 3.910 & -2.606 &
Mn\sci~ & 6440.971 &3.772 &-0.938 \\
Co\sci~ & 6411.884 & 2.542 & -2.228 & V\sci~  & 6431.623 & 1.950 & -0.287 &
Fe\sci~ & 6441.780 &5.033 &-1.928 \\
Ti\sci~ & 6412.192 & 3.581 & -1.480 & Pr\scii~ & 6431.805 & 1.416 & -0.339 &
Si\sci~ & 6442.777 &6.125 &-1.240 \\
Fe\sci~ & 6412.200 & 2.453 & -4.441 & Sm\scii~ & 6431.978 & 1.413 & -1.460 &
Fe\scii~ & 6442.955 &5.549 &-2.885 \\
Ti\sci~ & 6413.108 & 0.048 & -5.006 & Ni\sci~ & 6431.994 & 3.542 & -2.500 &
Fe\sci~ & 7507.261 &4.415 &-3.482 \\
Sc\sci~ & 6413.324 & 0.021 & -2.677 & Er\sci~ & 6432.531 & 1.309 & -0.919 &
Fe\sci~ & 7508.533 &4.988 &-2.318 \\
Ga\sci~ & 6413.468 & 3.073 & -0.300 & Nd\sci~ & 6432.650 & 1.287 & -0.160 &
Fe\sci~ & 7508.623 &4.143 &-2.741 \\
C\sci~  & 6413.550 & 8.771 & -1.380 & Fe\scii~ & 6432.680 & 2.891 & -3.540 &
Ti\sci~ & 7509.728 &1.749 &-3.030 \\
Pr\scii~ & 6413.678 & 1.132 & -0.799 & V\sci~  & 6433.162 & 1.945 & -1.308 &
Si\sci~ & 7510.785 &5.984 &-1.770 \\
Mn\sci~ & 6413.945 & 3.763 & -1.910 & Nb\sci~ & 6433.204 & 0.657 & -1.480 &
Fe\sci~ & 7511.020 &4.178 & 0.059 \\
Si\sci~ & 6414.188 & 6.083 & -2.450 & Si\sci~ & 6433.337 & 5.614 & -3.210 &
Nd\scii~ & 7511.125 &1.773 &-0.871 \\
Ni\sci~ & 6414.581 & 4.154 & -1.220 & Si\sci~ & 6433.457 & 5.964 & -1.590 &
Fe\sci~ & 7512.115 &2.279 &-5.195 \\
Ni\sci~ & 6414.796 & 4.165 & -2.463 & Fe\scii~ & 6433.814 & 6.219 & -2.371 &
Fe\sci~ & 7512.139 &4.143 &-2.947 \\
Si\sci~ & 6414.980 & 5.871 & -1.100 & Zr\sci~ & 6434.388 & 2.798 & -0.090 &
Cr\scii~ & 7512.745 &4.756 &-2.630 \\
Fe\sci~ & 6415.505 & 4.435 & -3.912 & Ce\sci~ & 6434.388 & 0.028 & -1.437 &
Nd\scii~ & 7513.736 &0.933 &-1.241 \\
S\sci~  & 6415.522 & 7.870 & -1.360 & Y\sci~  & 6435.004 & 0.066 & -0.820 &
Fe\scii~ & 7515.091 &5.823 &-2.362 \\
Nd\scii~ & 6425.779 & 1.649 & -0.935 & V\sci~  & 6435.158 & 1.942 & -1.458 &
Fe\scii~ & 7515.831 &3.903 &-3.432 \\
\end{tabular}
\end{center}
\label{lines}
\end{table*}
\renewcommand{\arraystretch}{1.0}

\section{Observations}

High resolution spectra of \object{HD~199178} were measured in July 1994, 
August 1994, November 1994 and July 1995. The July 1994 data set was 
obtained with the 2 m Ritchey-Chreti\'en telescope coud\'e-spectrograph of 
the National Astronomical Observatory (Rozhen), Bulgaria. The rest of the 
spectroscopic observations were made using the SOFIN high resolution \'echelle 
spectrograph  at the 2.5 m Nordic Optical Telescope (NOT), La Palma, Spain. 
The observed spectral regions were  $6409-6414$ {\AA} (August 1994 and July 
1995), $6427-6441 $ {\AA} (all seasons) and $7509-7514$ {\AA} (August 1994, 
November 1994 and July 1995). The typical signal-to-noise ratio was between
200 and 300. The spectral observations are summarized in Table~\ref{obs}.

The spectral observations were reduced with the 3A software system
(Ilyin \cite{ilyin}). The reduction included bias, cosmic ray, flat
field and scattered light correction, wavelength calibration and 
normalization. Special care was taken during the last two steps, where
errors can cause artifacts in the surface images. 

The calibration of the wavelength scale was obtained from Th-Ar comparison 
spectra.  For the \'echelle spectra, we used all available spectral orders 
of the comparison frames to calculate a two-dimensional ($k \lambda$) 
wavelength solution, taking into account distortions of the line positions 
(Ilyin \cite{ilyin}). Small shifts between the stellar and comparison 
spectra (e.g. due to flexure of the spectrograph) were corrected using 
atmospheric lines. The spectra were rebinned to a common wavelength grid.

The photospheric absorption lines of \object{HD~199178} are broad and it 
is hard to determine the real continuum in the spectrum. Therefore the 
continuum normalization had to be done in two steps. A ``quasi-continuum'' 
consisting of points near to the real continuum level was used as a first 
approximation. The final normalization was done by utilizing a rotationally 
broadened synthetic spectrum.

Simultaneous or nearly simultaneous Johnson $V$-photometry for July 1994,
August 1994 and July 1995 was obtained with the 60 cm telescope at the
Mount Maidanak Observatory and the APT Phoenix-10 telescope (subsets
SET=108, 109 and 111, Jetsu et al. \cite{photo}).

\section{Temperature maps}

The following three spectral regions were used for surface imaging: 
6409.0 -- 6413.5 \AA~ (strongest line Fe\sci~$\lambda 6411.649$ \AA), 
6427.5 -- 6441.0 \AA~ (Fe\sci~$\lambda 6430.846$ \AA; Fe\scii~$ \lambda
6432.680$ \AA; Ca\sci~$ \lambda 6439.075$ \AA) and 7509.0 -- 7513.5 \AA~ 
(Fe\sci~$ \lambda 7511.020$ \AA). The local line profiles were calculated for 
20 limb angles and temperatures 3500 -- 6500 K (with a step of 250 K) using a 
code written by Berdyugina (\cite{sloc}) and atmospheric models by
Kurucz (\cite{kurucz}) with solar element abundances. Spectral line parameters
were obtained from the Vienna Atomic Line Database (Kupka et al. \cite{vald2}).
All significant atomic lines and the most important molecular lines (CN and 
TiO) were included in the calculations. The final adjustment of the $gf$ 
values was done after analyzing the wavelength dependence of the deviation 
between a synthetic spectrum of \object{HD~199178} and the mean of the 
observations. Of the 111 lines used in the calculations, the $gf$
value of 36 lines was changed. The changes in the strongest
lines, Fe\sci~$\lambda 6411.649$ \AA, Fe\sci~$\lambda 6430.846$ \AA, 
Fe\scii~$ \lambda 6432.680$ \AA, Ca\sci~$ \lambda 6439.075$ \AA~ and 
Fe\sci~$ \lambda 7511.020$ \AA, were $-0.1 < \Delta \lg (gf) < 0.02$. It 
should be emphasized that the adopted $gf$ values may be biased by 
uncertainties in element abundances. A list of all atomic lines that we used, 
can be found in Table \ref{lines}.

\begin{table}[ht]
\caption{Stellar parameters used for the surface images in
Figs. \ref{jul94} -- \ref{jul95}}
\begin{center}
\begin{tabular}{lc}

\hline
Parameter & Adopted value \\
\hline
Gravity $\lg(g)$        &  3.5     \\
Inclination $i$ & $60^\circ$ \\
Rotation velocity \vsini          &  70 \kms \\
Differential rotation $\alpha$  & $-0.17$ \\
Rotation period $P_{\rm rot}$    &  3.3250   \\
Micro turbulence $v_{\rm micro}$ & 1.4 \kms \\
Macro turbulence $v_{\rm macro}$ & 4.0 \kms \\ 
Element abundances & solar \\

\hline
\end{tabular}
\end{center}
\label{spar}
\end{table}

The temperature maps were derived using the INVERS7-9PD code, 
originally written by N.E. Piskunov, with some changes made by T. Hackman.
We first tested different values of rotational velocity (68 \kms $ \leq $
\vsini $ \leq $ 74 \kms), microturbulence (1.2 \kms $ \leq v_{\rm micro} \leq $
1.8 \kms), inclination ($50^\circ \leq i \leq 85^\circ$) and surface 
differential rotation ($ -0.20 \leq \alpha \leq 0.20 $). Differential 
rotation was implemented by adjusting the angular rotational velocity at each 
latitude $b$ using the Maunder rotation formula

\begin{equation}
\Omega(b) = \Omega_{\rm equator} (1 - \alpha \sin^2 b).
\label{eq:maund}
\end{equation}

\noindent
The best solution (the smallest deviation between the spectroscopic
observations and the calculated line profiles) was obtained with the 
stellar parameters given in Table \ref{spar}. 

Before combining all available spectral regions, separate images were
calculated for each line blend in order to check for
inconsistencies. The individual maps for each spectral region gave the same
main features, but contained more artifacts (small high contrast
features, especially at low latitudes) than the maps obtained from the 
combination of all spectral regions.

\begin{figure}
\setlength{\unitlength}{1mm}
\begin{picture}(88,160)
\put(-10,80){\begin{picture}(0,0) \includegraphics{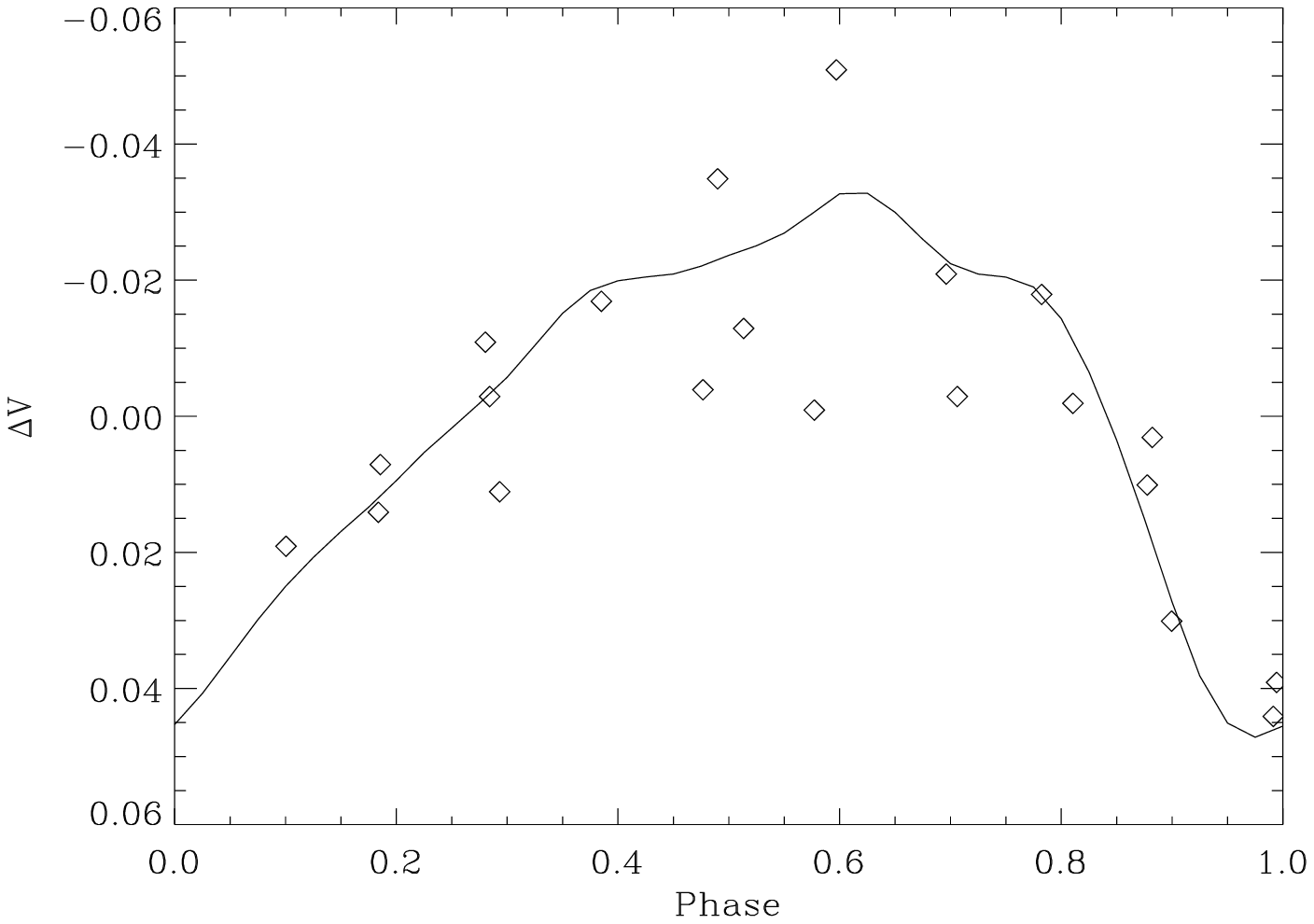} \end{picture}}
\put(-23,-20){\begin{picture}(0,0) \includegraphics{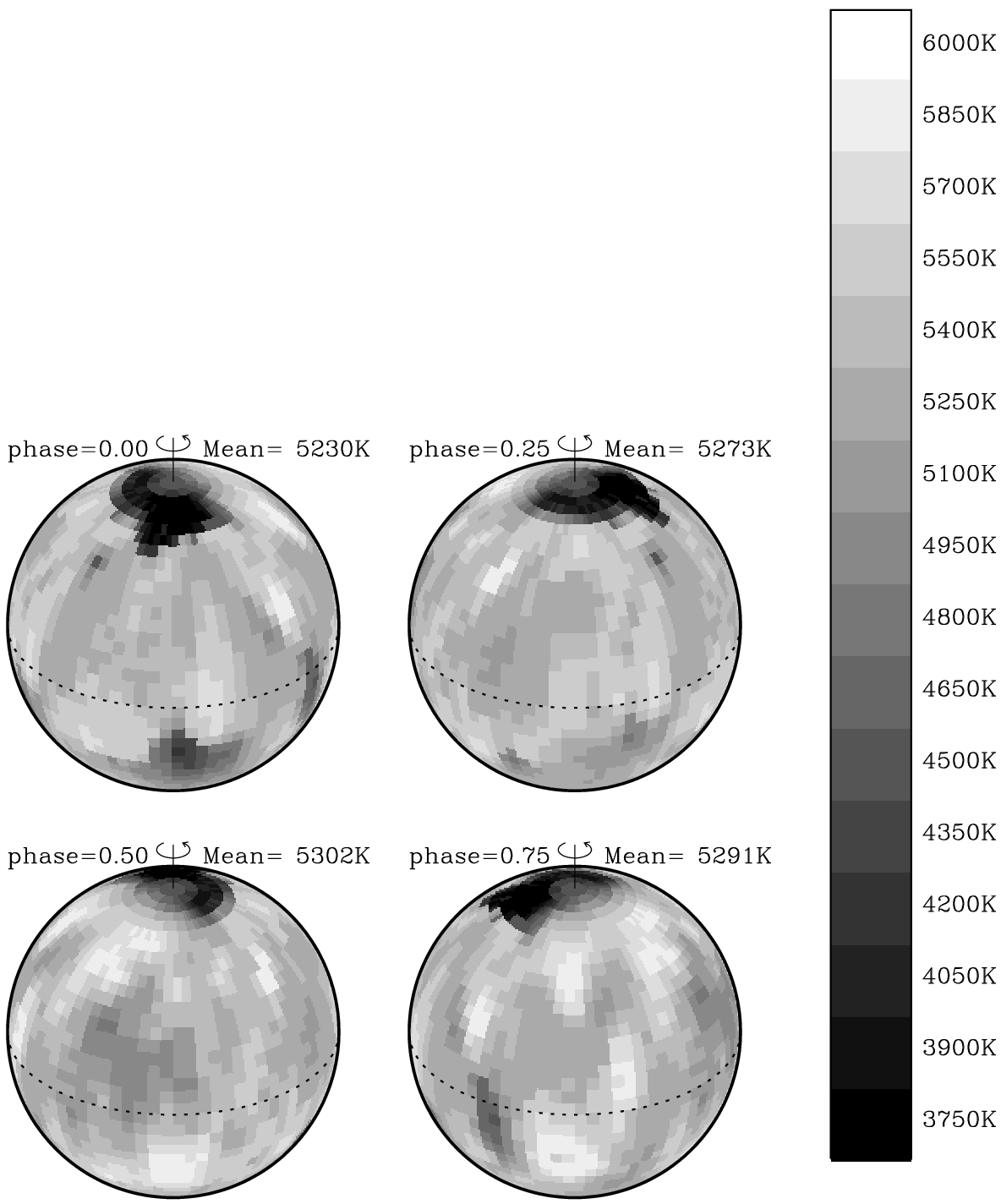} \end{picture}}
\put(5,-87){\begin{picture}(0,0) \includegraphics{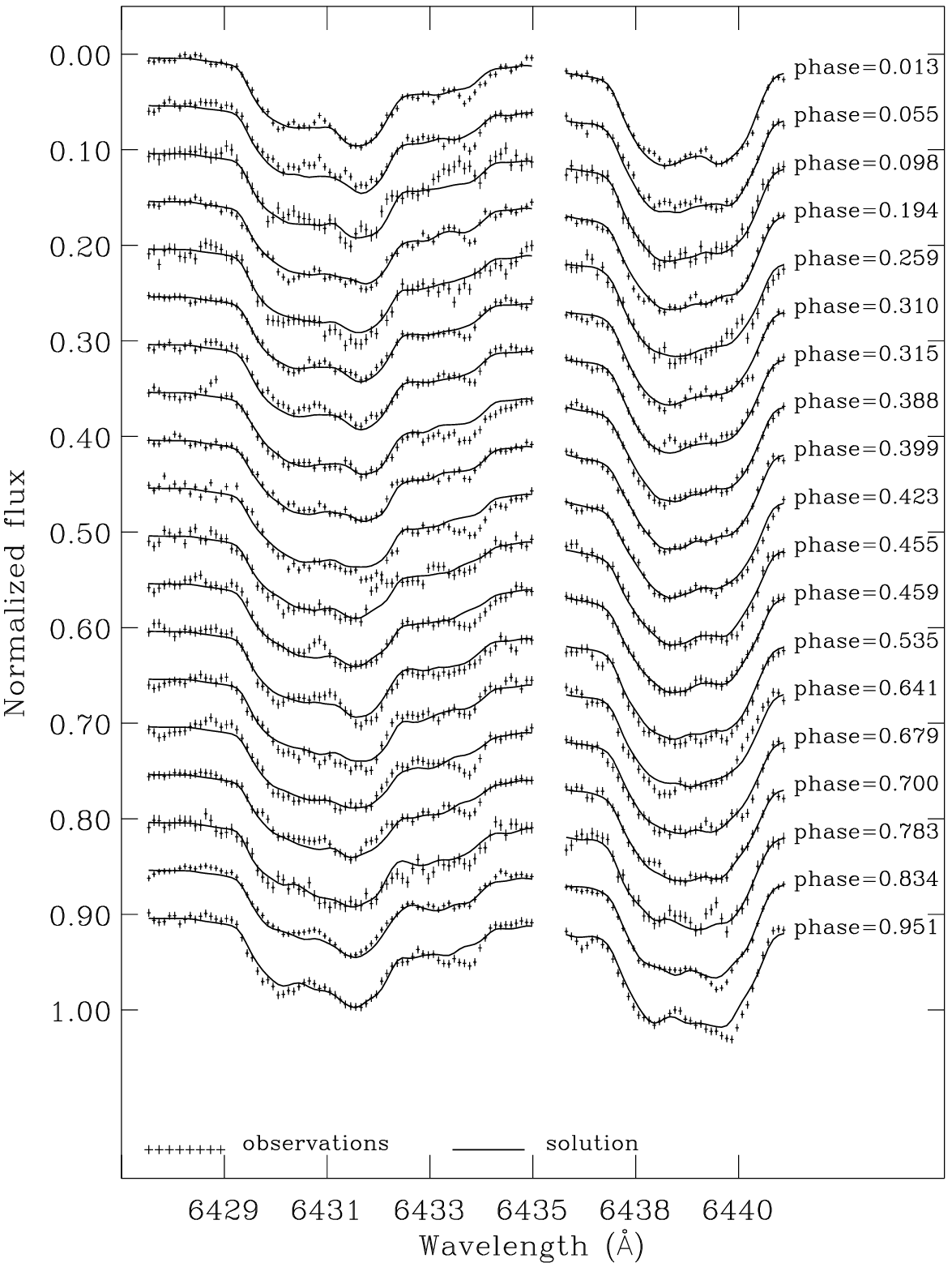} \end{picture}}
\end{picture}
\caption{The July 1994 image of \object{HD~199178} with calculated and
observed spectra and photometric $V$-curve. The mean deviation of the
spectroscopic observations from the model was $d=0.56\%$.}
\label{jul94}
\end{figure}

\begin{figure}
\setlength{\unitlength}{1mm}
\begin{picture}(88,130)
\put(-10,50){\begin{picture}(0,0) \includegraphics{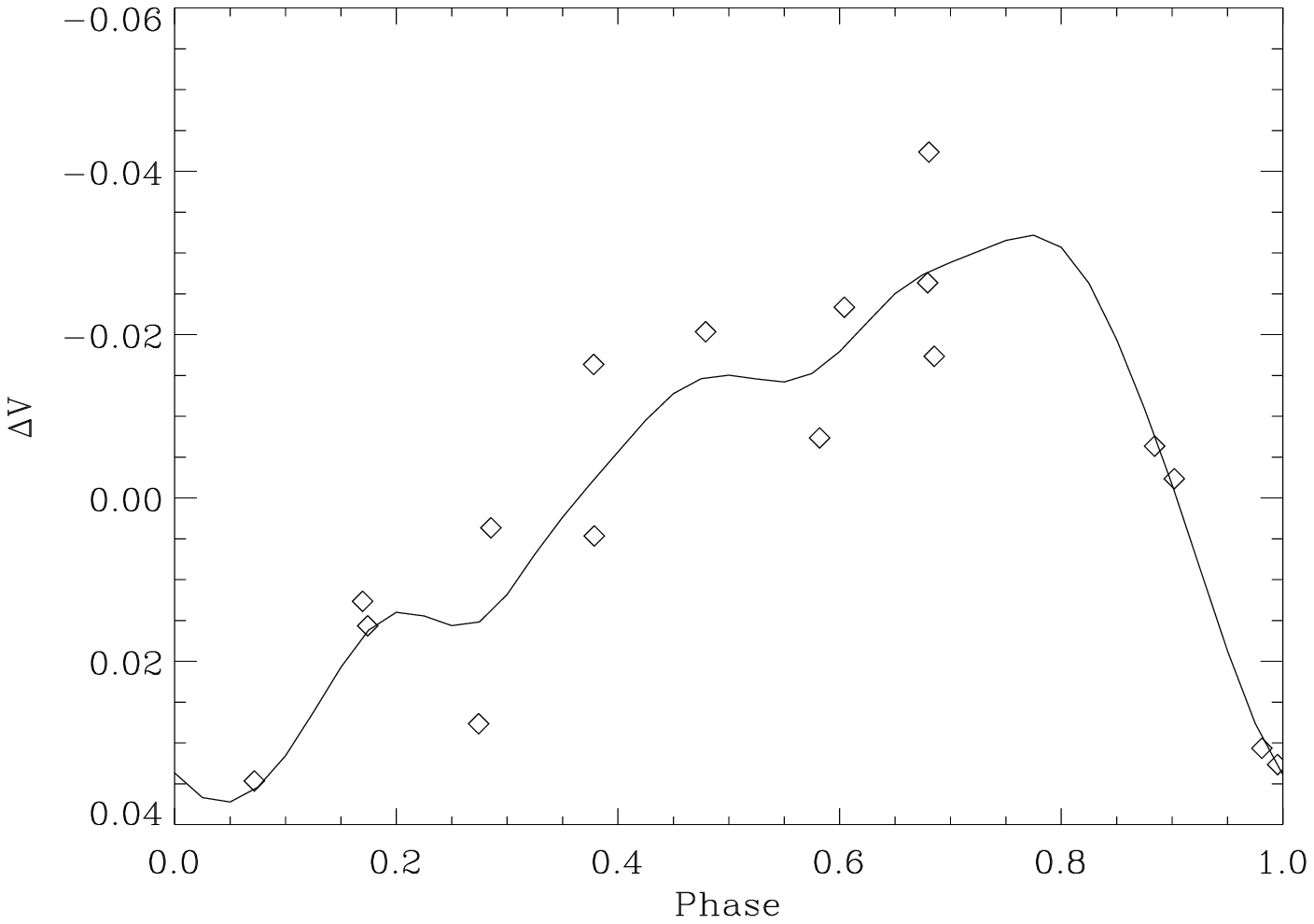} \end{picture}}
\put(-23,-50){\begin{picture}(0,0) \includegraphics{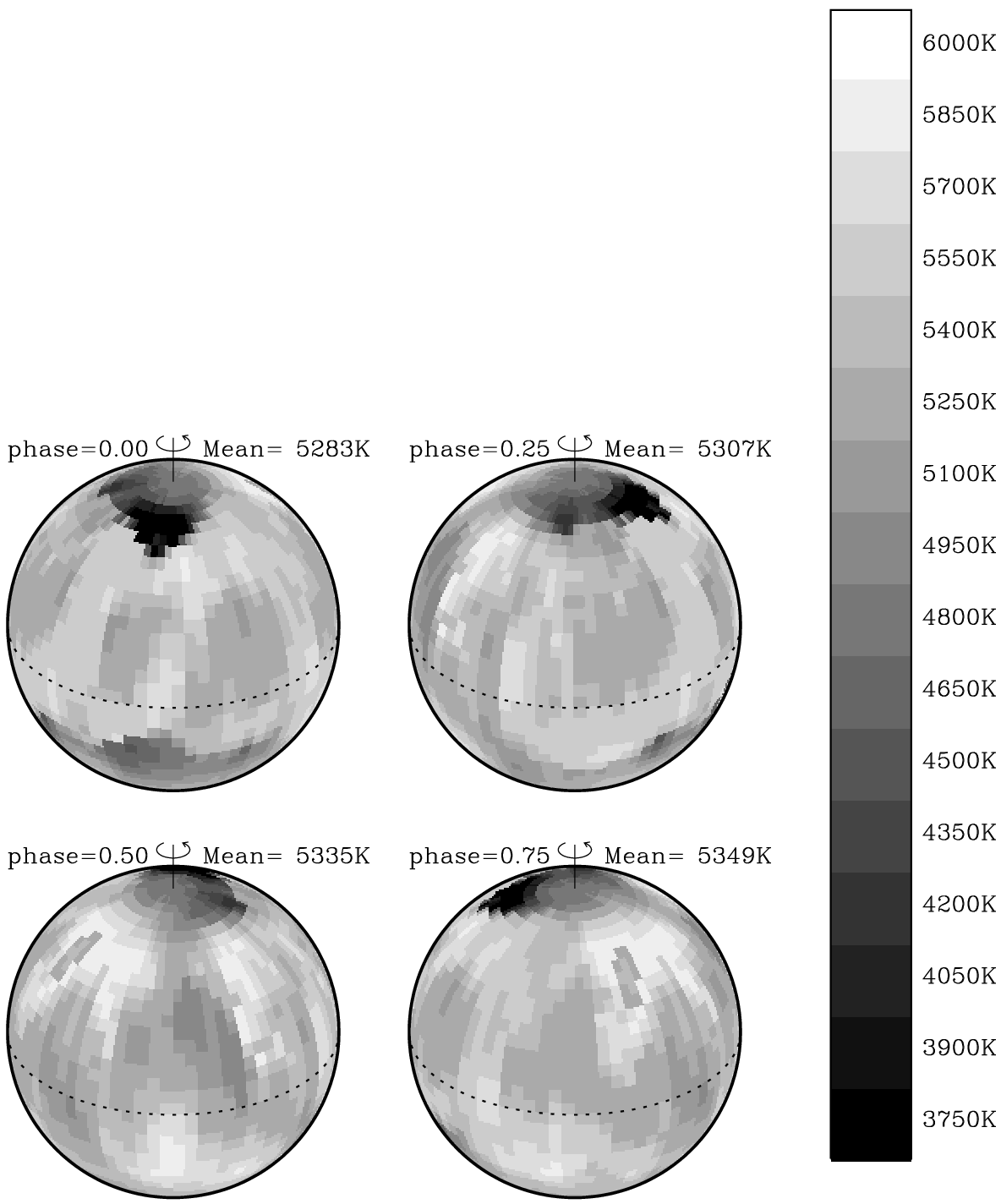} \end{picture}}
\put(-5,-92){\begin{picture}(0,0) \includegraphics{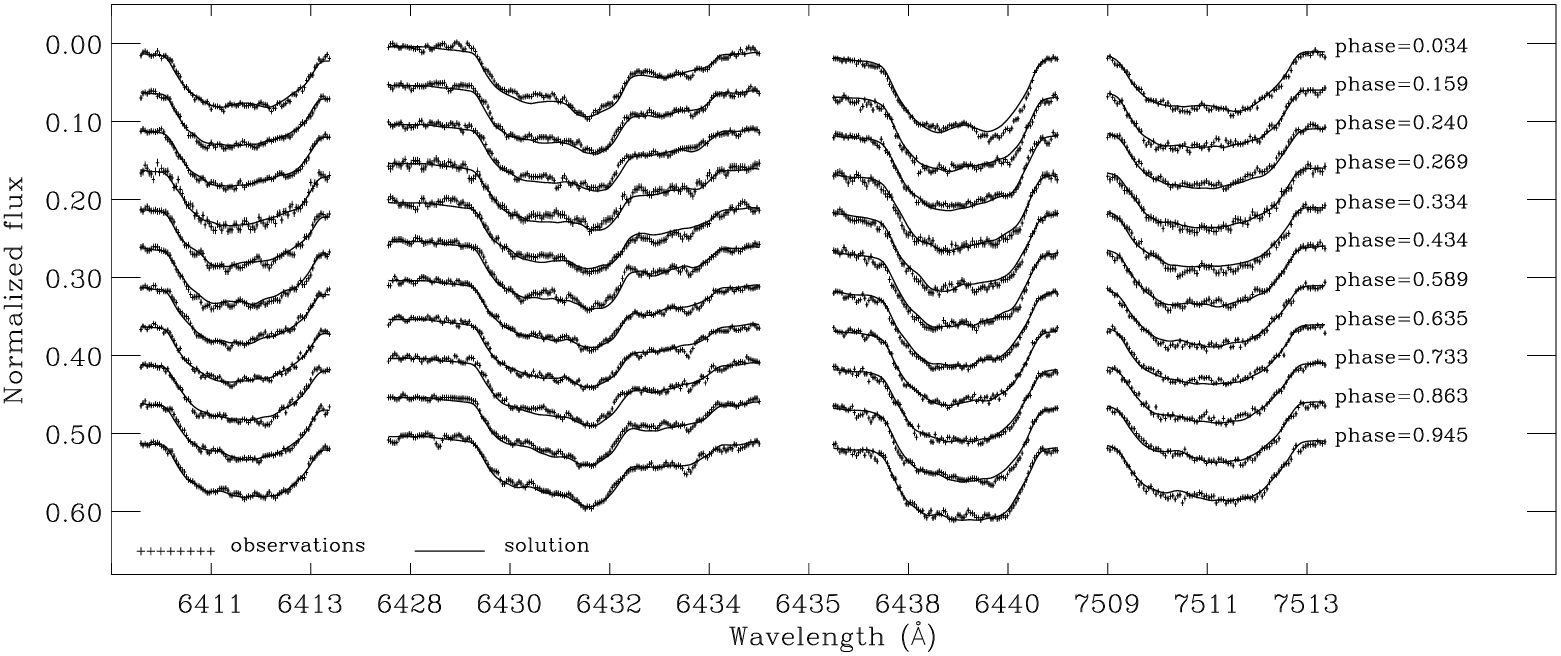} \end{picture}}
\end{picture}
\caption{The August 1994 image of \object{HD~199178} with calculated and
observed spectra and photometric $V$-curve. The mean deviation of the
spectroscopic observations from the model was $d=0.43\%$.}
\label{aug94}
\end{figure}

\begin{figure}
\setlength{\unitlength}{1mm}
\begin{picture}(0,100)    
\put(-8,-25){\begin{picture}(0,0) \includegraphics{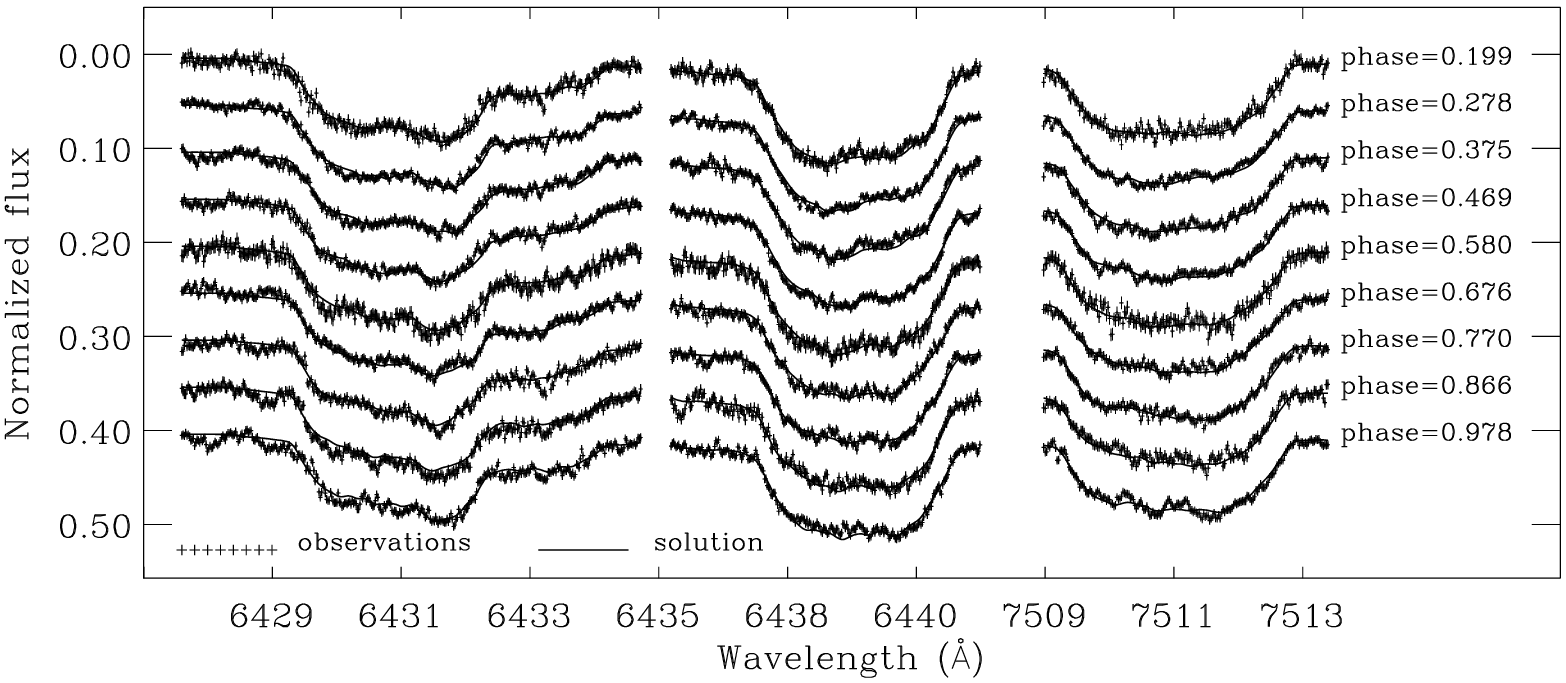} \end{picture}}
\put(-23,-80){\begin{picture}(0,0) \includegraphics{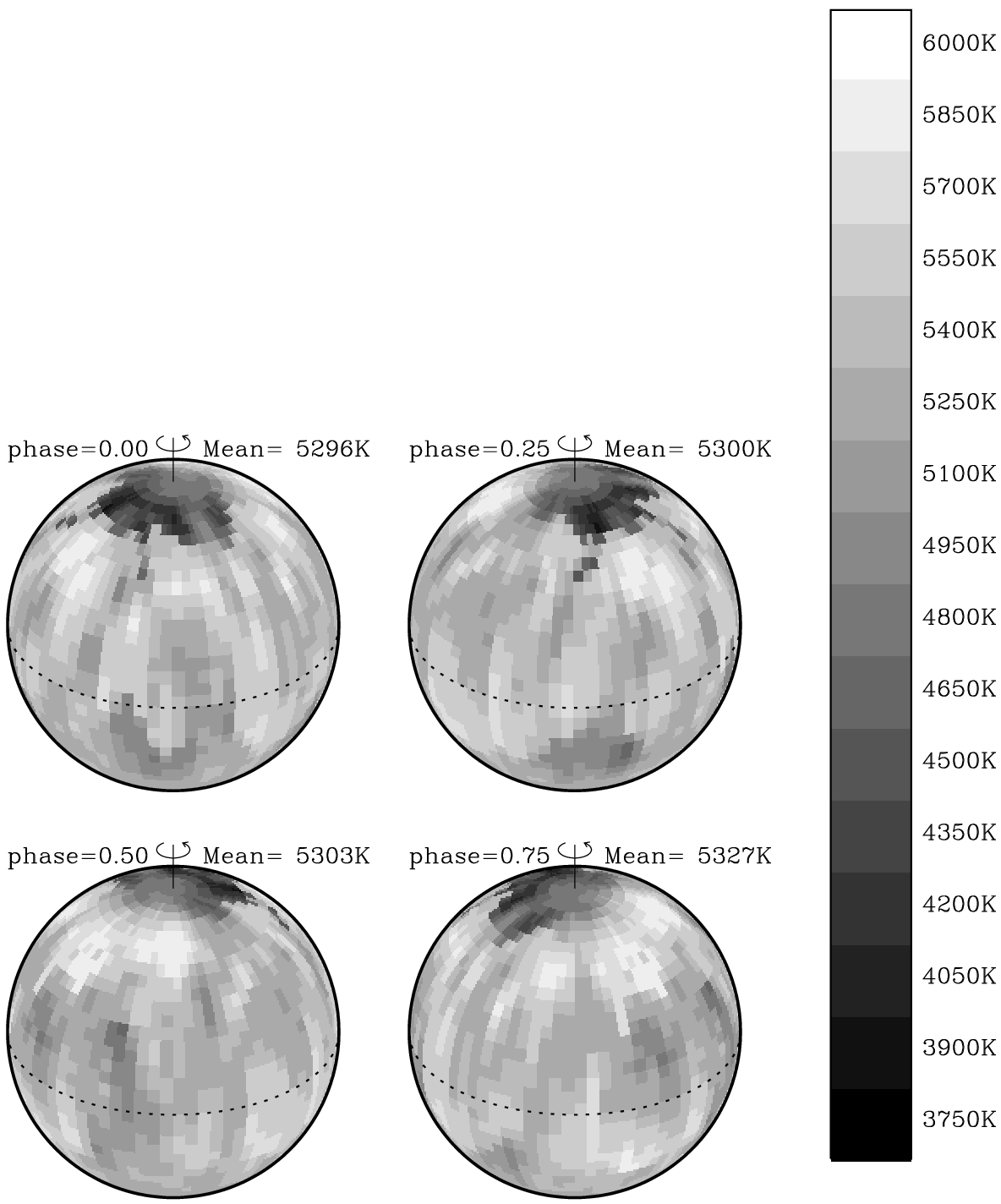} \end{picture}}
\end{picture}
\caption{The November 1994 image of \object{HD~199178} with calculated and
observed spectra (no $V$-photometry available). The mean deviation of the
spectroscopic observations from the model was $d=0.50\%$.}
\label{nov94}
\end{figure}

\begin{figure}
\setlength{\unitlength}{1mm}
\begin{picture}(88,130)
\put(-10,50){\begin{picture}(0,0) \includegraphics{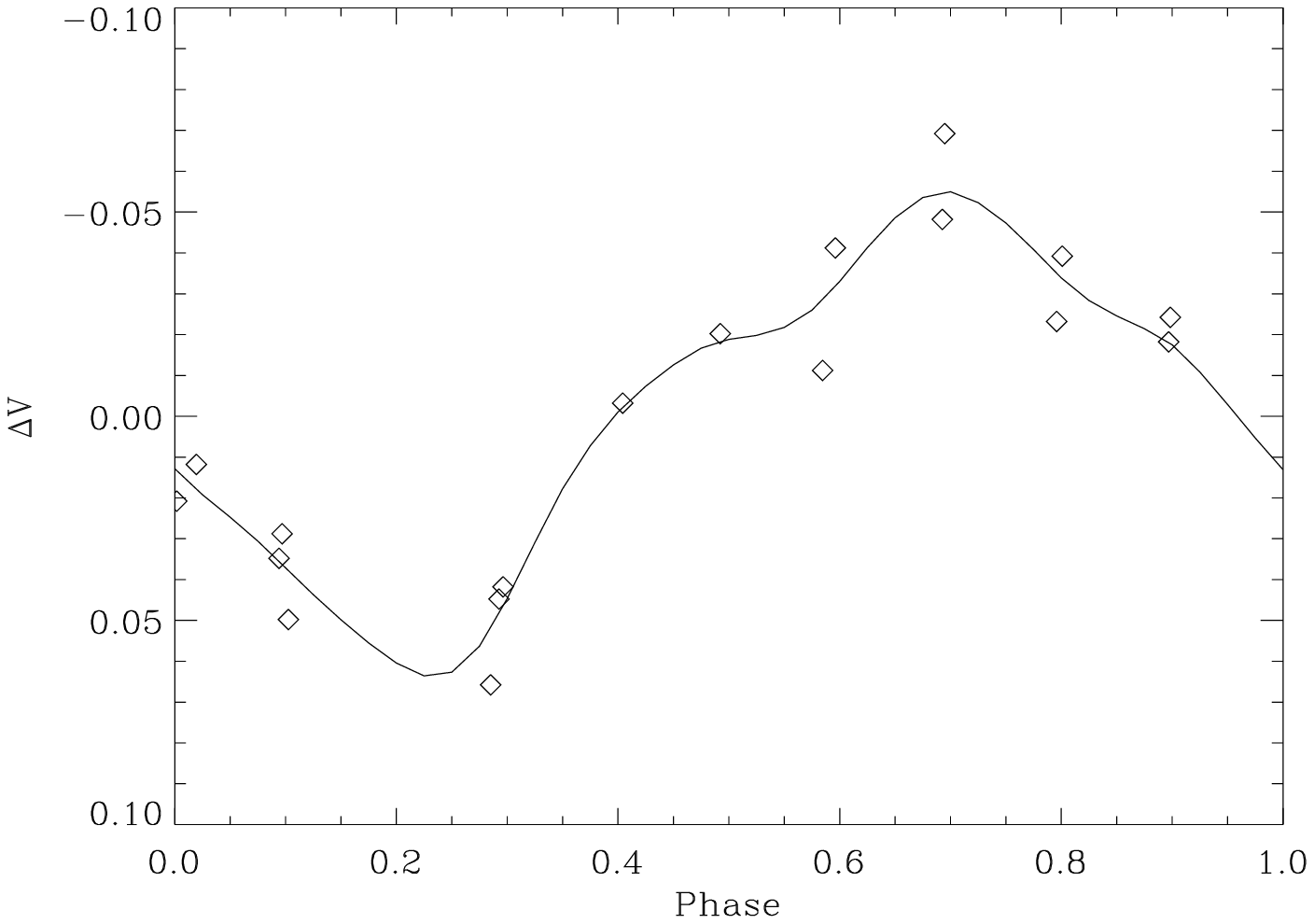} \end{picture}}
\put(-23,-50){\begin{picture}(0,0) \includegraphics{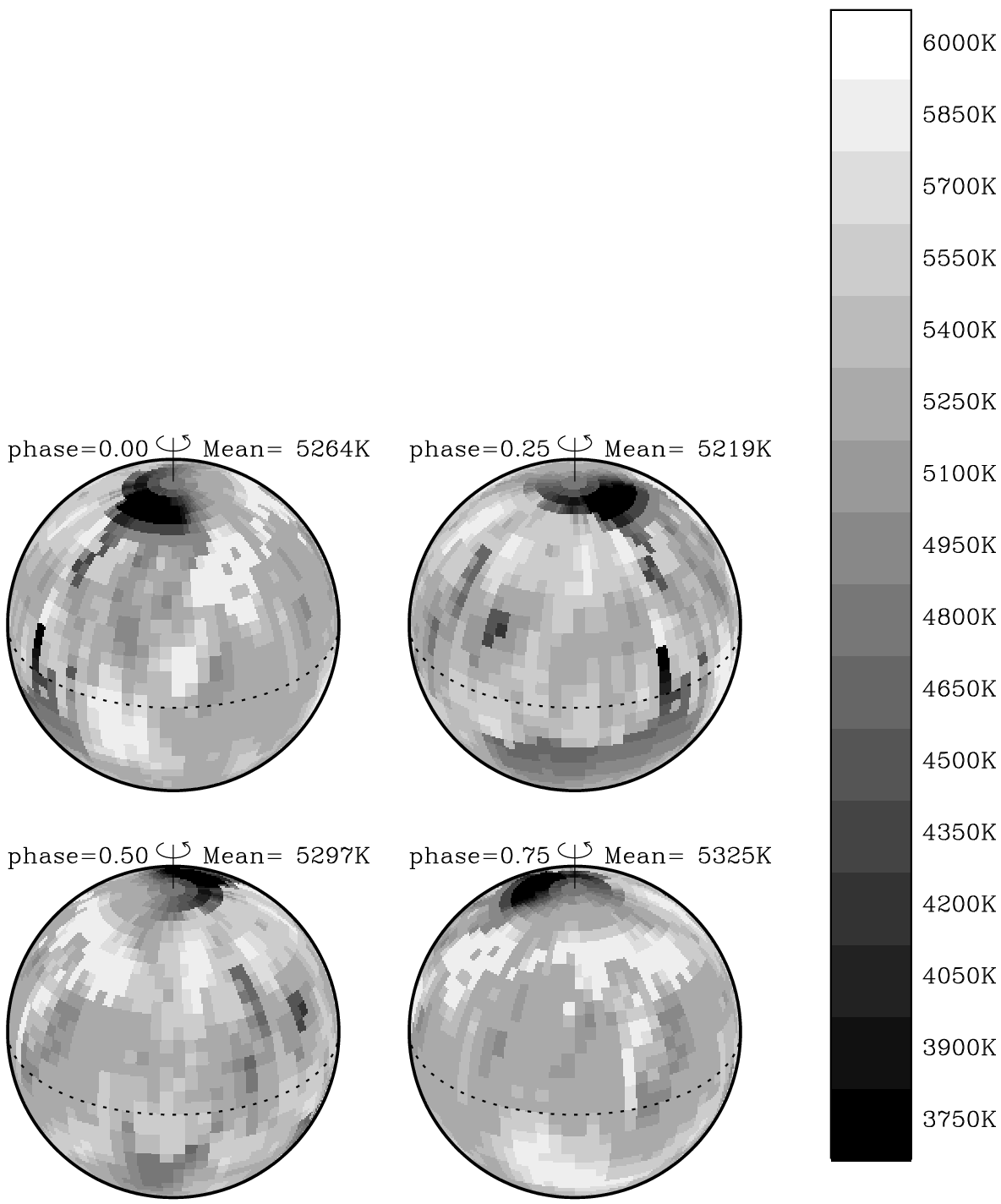} \end{picture}}
\put(-5,-92){\begin{picture}(0,0) \includegraphics{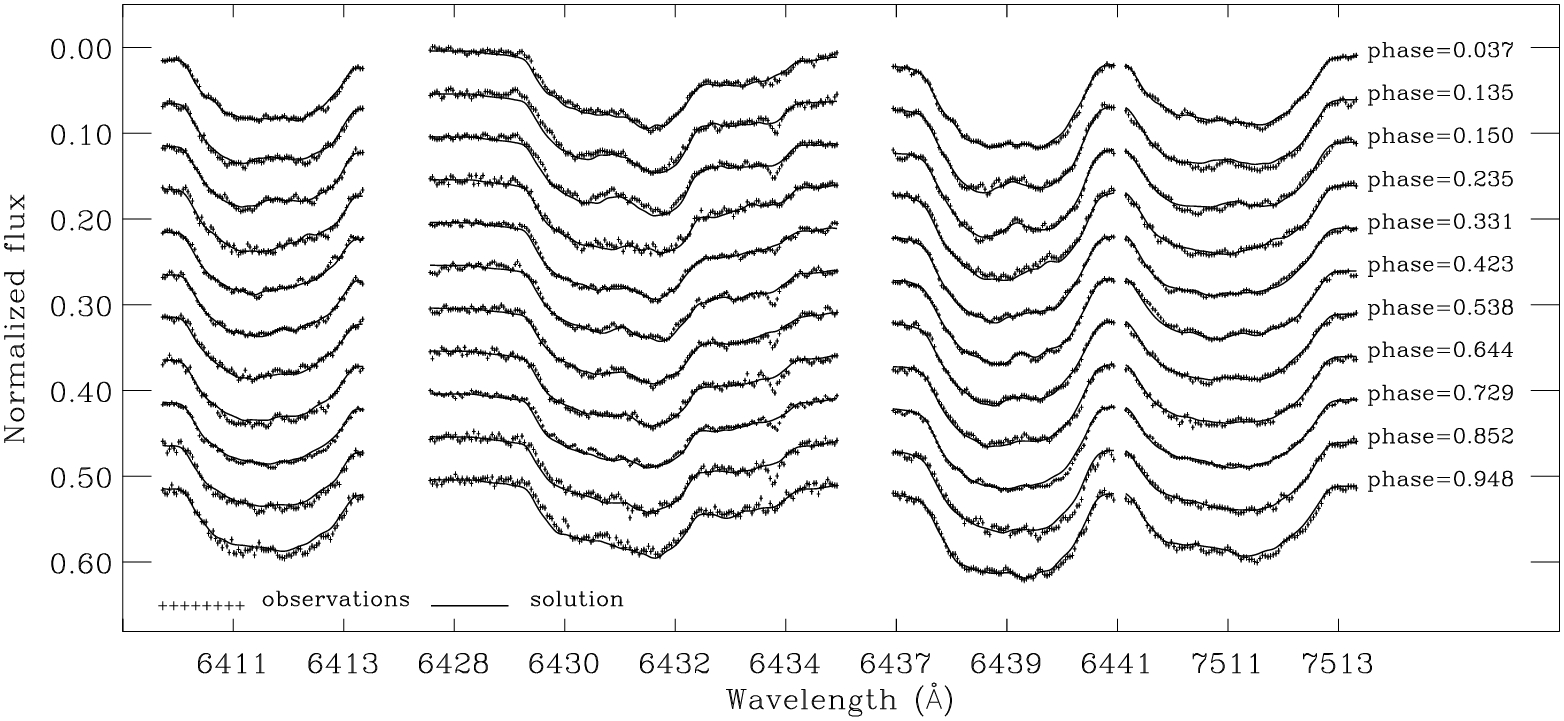} \end{picture}}
\end{picture}
\caption{The July 1995 image of \object{HD~199178} with calculated and
observed spectra and photometric $V$-curve. The mean deviation of the
spectroscopic observations from the model was $d=0.37\%$.}
\label{jul95}
\end{figure}

\begin{figure}
\setlength{\unitlength}{1mm}
\begin{picture}(88,110)
\put(-23,-150){\begin{picture}(0,0) \includegraphics{h1748f5a.ps} \end{picture}}
\put(-23,-185){\begin{picture}(0,0) \includegraphics{h1748f5b.ps} \end{picture}}
\put(-23,-220){\begin{picture}(0,0) \includegraphics{h1748f5c.ps} \end{picture}}
\end{picture}
\caption{The July 1995 image with differential rotation coefficients $\alpha =
-0.17$ (same as Fig. \ref{jul95}), 0.0 and 0.15.} 
\label{dr}
\end{figure}

All surface images show a large cool spot remaining at approximately the same
latitude (Figs. \ref{jul94}--\ref{jul95} and Table \ref{spots}). The mean 
temperature of the visible surface is 5300 K. The polar region is about 
4900 K. The lower latitude features in the images are probably artifacts. In
all maps there are symmetry effects typical for surface imaging of
stars with relatively high inclination ($i \ge 60^o$), i.e. ``shadows'' of the 
high latitude spot below the equator. The bright features, reaching a maximum 
of $T_{\rm eff} \approx$ 5800 K, are probably also artifacts. 

The {\it July 1994} image shows a large high latitude cool spot (3700 K) with a
longitudinal appendage (3900 K). In the {\it August 1994} image, the coolest 
part of the spot (3700 K) is more concentrated and the appendage (4200 K) is
weaker. In {\it November 1994}, the spot appears to be partly disintegrated. 
The coolest part of the spot (4100 K) seems to be at the position of the 
previous appendage. No photometric data was available for the November 1994 
image, but there were clear changes in the light curves for July --
September 1994 (subsets SET=108--110, Jetsu et al. \cite{time}), which
support the spot evolution. The {\it July 1995} image resembles the
July 1994 image, but with an additional new appendage stretched ``southwards''.
The coolest part of the spot (3700 K) has regained its strength.

\begin{table}[ht]
\caption{Temperatures $T_{\rm spot}$, longitudes $\ell$ and latitudes
$b$ of the coolest concentrated parts of the spot in each
image, and photometric periods $P$ for each set from Jetsu et al. \cite{time}.
(The photometric periods are listed only for comparison, each image was 
derived using the same photometric period $P=3.3250$, i.e. the mean period for
the years 1994-1995.) At phase $\phi$ the longitude $\ell = - 360 ^{\circ} 
\phi$ is at the center of the visible stellar disk}
\begin{center}
\begin{tabular}{lcccl}

\hline

Image & $T_{\rm spot}$ (K) & $\ell$ ($^{\circ}$) & $b$ ($^{\circ}$) & $P$ (d)\\

\hline

July 1994     & 3700 & -5   & 67 & 3.352  \\
August 1994   & 3700 & -15  & 61 & 3.384  \\
November 1994 & 4100 & -70  & 65 &  --    \\
July 1995     & 3700 & -39  & 68 & 3.2829 \\

\hline
\end{tabular}
\end{center}
\label{spots}
\end{table}

The best surface imaging solution was achieved with the differential rotation 
coefficient $\alpha=-0.17 $. This implies strong antisolar differential 
rotation in \object{HD~199178}, i.e. that the pole would rotate faster than 
the equator. It seems that current models for global circulation in outer 
stellar convection zones predict only solar-like differential rotation (cf. 
Kitchatinov \& R\"udiger \cite{kitch}). Therefore we also wanted to test the 
possibility of no differential rotation and solar-like differential rotation.
Our Fig. \ref{dr} displays the July 1995 surface imaging solutions obtained
with $\alpha=0$ (rigid body rotation) and $\alpha = 0.15$ (solar-like 
differential rotation). The best fits with $\alpha=0$ and $\alpha=0.15$ were 
achieved by increasing $v\sin i$ to 72 {\kms} and 74 {\kms}, respectively. The
mean deviation ($d$) of the calculated line profiles from the observations was
slightly higher for $\alpha = 0$ and $\alpha =0.15$ than for $\alpha
=-0.17$, i.e. the antisolar case provided a better fit to
the data (Table \ref{c95}). 

The images in Fig. \ref{dr} demonstrate the difficulties in separating
the effect of a polar spot from that of differential rotation.
With no differential rotation the polar region is
$\sim$ 800 K cooler than the mean surface temperature, while 
$\alpha = 0.15$ gives a cool polar spot with $T_{\rm eff} \approx $
3900 K, much like the results by Strassmeier et al. (\cite{klasu2}). 
The same general effect was verified for all other seasonal images. Except for
the polar cap, the images in Fig. \ref{dr} are very similar. The
coolest part of the spot remains at approximately the same position. Similar
artifacts can also be seen in all three images, in particular some
bright features and smaller spots near the equator.

\begin{table}[ht]
\caption{The quality of the July 1995 surface imaging solution
using different values of $\alpha$: mean deviation $d$ of the calculated
line profiles from the spectroscopic observations and linear
correlation coefficient $r$ between next-to-neighbouring residuals}

\begin{center}

\begin{tabular}{crcc}

\hline

\vsini   & $\alpha$ & $d$      & $r$    \\

\hline

70 \kms  & -0.17    & 0.373\%  &  0.507 \\
72 \kms  & 0.00     & 0.388\%  &  0.540 \\
74 \kms  & 0.15     & 0.410\%  &  0.572 \\

\hline
\end{tabular}
\end{center}
\label{c95}
\end{table}

\section{Differential rotation or polar spot?}

The surface imaging solution with antisolar surface differential rotation might
raise the question of whether this is a real phenomenon or just a trick
to eliminate a cool polar spot. Thus, what we need is a way to determine which 
one, the polar spot or differential rotation, gives a better solution.
When analyzing the T Tauri star \object{Sz 68}, Johns-Krull \& Hatzes 
(\cite{sz68}) tackled this problem by calculating separate temperature maps 
for different spectral lines. They argued for solid body rotation and a polar 
spot, because this alternative gave a larger correlation between the 
individual maps. 

We tested whether this approach works in our case.
First we {\it simulated} synthetic spectroscopic ``observations''
using as input a stellar surface with a single high latitude spot, but
{\it no polar spot}, and stellar parameters \vsini=70 \kms, $\alpha =
-0.2$, $i = 60^{\circ}$ (CASE \sci). To imitate real observations, we added
normally distributed noise ($\sigma = 0.30\%$) to the synthetic observations. 
These synthetic spectra were then used to calculate temperature maps with two 
sets of stellar parameters:  \vsini=70 \kms, $\alpha = -0.2$, $i = 60^{\circ}$
and \vsini=72 \kms, $\alpha = 0.0$, $i = 60^{\circ}$. Maps were calculated both
using all spectral regions and separately for each spectral region
(6411\AA, 6431 \AA, 6439 \AA~ and 7511 \AA). With the correct stellar 
parameters, we retrieved a surface imaging solution consistent with the input 
stellar surface. As expected, neglecting the differential rotation, i.e. using 
$\alpha = 0.0$, resulted in a solution with a cool polar spot in
addition to the high latitude spot.

For both sets of stellar parameters we calculated the correlations between 
the maps retrieved for the different spectral lines. The test results showed 
that the correlation was higher when the differential rotation was wrongly 
neglected (Tables \ref{corr1} and \ref{corr2}). The reason for this is clear. 
Neglecting a substantial differential rotation, one always obtains the same 
artifact, a polar spot, regardless of the spectral line used for the image. 
The presence of the same polar spot in all images will naturally increase the 
correlation between the images. The correlation only measures the consistency 
between the calculated images. It will not reveal the inconsistency between the
original data (no polar spot) and the solution (polar spot).

\begin{table}[ht]
\caption{CASE \sci: Correlation between solutions for different
spectral lines. The solutions were retrieved {\it neglecting} differential
rotation ($\alpha=0$), although the original data was simulated with a 
differential rotation coefficient $\alpha=-0.2$ } 

\begin{center}

\begin{tabular}{lllll}

\hline

         & 6411 \AA & 6431 \AA & 6439 \AA & 7511 \AA \\

\hline

6411 \AA & r=1.000  &          &          &          \\
6431 \AA & r=0.975  & r=1.000  &          &          \\ 
6439 \AA & r=0.959  & r=0.967  & r=1.000  &          \\
7511 \AA & r=0.961  & r=0.953  & r=0.970  & r=1.000  \\

\hline
\end{tabular}
\end{center}
\label{corr1}
\end{table}

\begin{table}[ht]
\caption{CASE \sci: Correlation between solutions for different
spectral lines. Both the solutions and the original data were
simulated {\it with} the differential rotation coefficient $\alpha=-0.2$}

\begin{center}

\begin{tabular}{lllll}

\hline

         & 6411 \AA & 6431 \AA & 6439 \AA & 7511 \AA \\

\hline

6411 \AA & r=1.000  &          &          &          \\
6431 \AA & r=0.899  & r=1.000  &          &          \\ 
6439 \AA & r=0.918  & r=0.907  & r=1.000  &          \\
7511 \AA & r=0.887  & r=0.913  & r=0.892  & r=1.000  \\

\hline
\end{tabular}
\end{center}
\label{corr2}
\end{table}

An alternative approach to check the reliability of Doppler imaging maps, is
to estimate the systematic errors in the calculated line profiles. As opposed 
to noise, the systematic errors of neighbouring wavelength points will
correlate. In other words, systematic errors will cause the neighbouring data 
points to tend to be on the same side of the calculated curve. Not 
surprisingly, the linear correlation coefficient $r$ of neighbouring 
residuals, i.e. the systematic error, was significantly larger for the 
solution with the (false) solid body rotation than for the one with the 
(correct) differential rotation (Table \ref{testr}: 1st and 2nd line). The 
systematic error could also be seen in the slightly larger deviation $d$.

To show that this criterion (lowest correlation of residuals) does not 
favour any particular spot distribution, we also simulated synthetic 
observations {\it with a polar spot} and no differential rotation (CASE II). 
Again the simulated data was used to retrieve temperature maps with and 
without differential rotation. The correct stellar parameters gave a solution 
consistent with the input map. In the solution with artificial differential 
rotation the polar spot was reduced, as expected. But also in this case, an 
erroneous differential rotation coefficient caused correlation in the 
residuals and increased the deviation (Table \ref{testr}: 3rd and 4th line).

\begin{table}[ht]

\caption{The quality of the surface imaging solution for rigid body
($\alpha=0$) and antisolar differential rotation ($\alpha=-0.2$) in CASES I 
and II: mean deviation $d$ of the solution from the synthetic spectroscopic
``observations'' and linear correlation $r$ of subsequent residuals of
the line profiles}

\begin{center}
\addtolength{\tabcolsep}{-0.1cm}
\begin{tabular}{lcr|crcc}

\hline

\multicolumn{3}{c|}{Input parameters} & 
\multicolumn{2}{l}{Solution} & & \\

polar spot & \vsini & $\alpha$ & \vsini & $\alpha$ & $d$ & $r$    \\

\hline

no (CASE I) & 70       & -0.2    & 70       & -0.2     & 0.30 \%    &  0.035 \\
no (CASE I) & 70       & -0.2    & 72       &  0.0     & 0.33 \%    &  0.198 \\
\hline
yes (CASE II) & 72      &  0.0    & 70       & -0.2    & 0.32 \%    &  0.110 \\
yes (CASE II) & 72      &  0.0    & 72       &  0.0    & 0.30 \%    & -0.001 \\
\hline
\addtolength{\tabcolsep}{+0.1cm} 
\end{tabular}
\end{center}
\label{testr}
\end{table}

We conclude, that aiming for a maximum correlation between images obtained 
from different spectral lines will favour solutions with polar spots.
Furthermore, systematical errors in the calculated lines profiles increase, if
the value of $\alpha$ is erroneous.

In the simulated tests we correlated the errors of neighbouring
points. But in our real data of \object{HD~199178} neighbouring residuals will 
anyway be correlated, because the spectra have been rebinned to a common 
wavelength scale. This effect can be avoided by comparing each point in
the spectra to the point {\it two} steps ahead. The correlation of 
next-to-neighbouring residuals was checked for the July 1995 surface imaging 
solutions with the differential coefficients $\alpha=-0.17$, $\alpha=0.0$ and 
$\alpha=0.15$ (Fig. \ref{dr}). 

The results in Table \ref{c95} showed that the solution with $\alpha=-0.17$ 
gave both the best fit (lowest $d$) and smallest correlation of the residuals 
(lowest r). However, we note that the correlation was highly significant even 
for the best stellar parameters (\vsini=70 \kms, $\alpha=-0.17$, 
$i=60^{\circ}$). Apparently there still remained systematic errors in the 
solution. This could also be seen from the fact that the deviation was larger 
than the observational noise. The most likely cause for these systematic 
errors were inaccuracies in the calculated line profiles (missing lines, 
errors in spectral parameters or element abundances, non-LTE effects, etc.).
Even though all systematic errors could not be eliminated, we
consider negative surface differential rotation more likely than the
polar spot, because the best quality of the surface imaging
solution was achieved with $\alpha=-0.17$. Still, we emphasize that 
the $\alpha$-value was used to adjust the calculated profiles to fit the 
observations. Therefore our estimate for $\alpha$ may be biased by errors in 
the line calculations.

The presence of differential rotation is supported by the variation of the 
photometric period, although photometry alone can only give a lower limit for 
$\vert \alpha \vert$. Inserting the means of the spot latitudes and 
photometric periods of Table \ref{spots} into Eq. \ref{eq:maund}, the value of 
$\alpha=-0.17$ would give the rotation periods at the equator $P_{\rm equator}
\approx 3\fd81$ and the pole $P_{\rm pole} \approx 3\fd25$. The reliable 
photometric periods of \object{HD~199178} determined by Jetsu et al. 
(\cite{time}) were within the range $3\fd191 \pm 0.017 \le P_{\rm phot} \le 
3\fd559 \pm 0.017$. Since \object{HD~199178} is a rapid rotator, one would 
expected it to have only high latitude spots (Sch\"ussler \& Solanki 
\cite{schuss}). Therefore it is natural, that we do not find the equatorial 
rotation period from photometry. But the fact that our $P_{\rm pole}$ is 
larger than the smallest measured $P_{\rm phot}$ indicates a slight 
discrepancy.

By combining the time series analysis of photometry with individual surface 
images we can study if the photometric rotation period depends on the latitude 
of the main spot. Antisolar differential rotation would be seen as a shorter
period for higher latitude spots. We note that this is indeed the
case for the photometric periods and latitudes of the spot
in Table \ref{spots}. But since we have only three measurements,
this tendency is far from statistically significant. Furthermore,
the differences in the spot latitudes obtained from the surface images
are too small to draw reliable conclusions.

There are other methods to measure the surface differential rotation. The 
approach most commonly used in Doppler imaging, is to cross-correlate constant 
latitude slices of images and calculate the shift in longitude as a function 
of the latitude (cf. Donati \& Collier Cameron \cite{abdor}). This approach 
requires that there really are long-lived spots at different latitudes. The 
method may also fail if there are too many artifacts in the
images. Artifacts will often appear at nearly the same longitude as the real
features or as vertical stripes in the images. A cross-correlation analysis 
of such images will give a nearly rigid body rotation. In our images, there 
is only one major spot, and the low latitude features are probably artifacts. 
For this reason, we do not attempt to cross-correlate our images.

Antisolar differential rotation has previously been found in some
active late-type stars. Johns-Krull (\cite{ttau}) analyzed the
rotationally broadened spectral line profiles of three T Tauri stars
and got the best fits with antisolar differential rotation. But he
also noted that a polar spot could produce this
effect. Weak antisolar differential rotation has been found in
Doppler images of the RS CVn stars UX Ari (Vogt \& Hatzes
\cite{uxari}), HR 1099 (Vogt et al. \cite{hr1099}) and HD 106225 
(Strassmeier \cite{hd106225}; Hatzes \cite{hatzes}).

\section{Conclusions}

The surface images of \object{HD~199178} presented here resemble the results 
recently obtained for \object{FK Comae Berenices} (Korhonen et al. 
\cite{heidi} \& \cite{heidi2}). The dominating feature is a long-lived high 
latitude spot (or spot group), with slightly changing position, shape and 
strength. Such changes could already be seen in the July 1994 and August 1994 
images. In the November 1994 image, the spot had weakened and partly 
disintegrated. In July 1995, the spot had recovered its strength. The changes 
in the photometric light curves of \object{HD~199178} during 1994 and 1995 can 
thus be explained by evolution of a {\it single} large spot rather than by 
abrupt changes of the whole spot configuration.

The long-lived spot structure suggests there is a stable active
longitude in \object{HD~199178}. The high latitude of the spot could be
explained by that the Coriolis force dominates over the buoyancy force
in the magnetic flux tubes  erupting from the deep parts of the
convection zone (Sch\"ussler \& Solanki \cite{schuss}).  The changes
in the spot latitude could then be connected to variations in the magnetic
field strength. The changes in the latitude will in turn lead to drifts
in the longitude if there is surface differential rotation.

In previous temperature maps of \object{HD~199178}, the ``flat bottomed''
absorption line profiles have  been interpreted as a signature of a large and
very cool polar spot (Vogt \cite{vogt1}; Strassmeier et al. \cite{klasu2}). We
achieved the best fit to the spectral observations by introducing an 
antisolar surface differential rotation of $\alpha=-0.17$. With
$\alpha \ge 0$, a polar spot appeared in the images. We
considered the solution with antisolar differential
rotation to be more likely than the polar spot, because it reduced both
the discrepancy between the observed and calculated line profiles, as
well as the correlation between residuals of neighbouring wavelength points
(i.e. systematical errors in the calculated line profiles). 

During 1994 and 1995 the photometric rotation period varied by 3 \%. Jetsu et 
al. (\cite{time}) interpreted these variations as a signature of surface 
differential rotation. Our images show that the photometric minimum of the 
light curves was apparently caused by the same persistent high latitude spot 
(group). Therefore the changes in the photometric period could not be explained
by a changing pattern of spots. We also note that in our surface images the 
changes in the spot latitude were less than 10 degrees. The differential 
rotation would have to be strong in order to cause the observed changes in the
photometric rotation period.

There are certainly other alternative explanations for the flat bottomed line
profile shape than antisolar differential rotation. The very rapid rotation of 
FK Comae-type stars, perhaps being recently coalesced binaries, suggests 
there could be significant deviations from a spherical shape (cf. Weltey \& 
Ramsey \cite{nonsph}). Our understanding of absorption line formation in the
photosphere of these exceptional stars may also be insufficient. In particular,
the strong absorption lines may be subject to non-LTE effects. On the other 
hand, Bruls et al. (\cite{bruls}) showed that chromospheric activity and 
non-LTE effects could only explain the box-like shape of {\it some} (e.g. 
Fe \sci~ $\lambda$ 6430 \AA), but {\it not all} (e.g. Fe \sci~ $\lambda$ 6411
\AA~ and Ca \sci~ $\lambda$ 6439 \AA) of the lines used in our analysis. 

Weak antisolar differential rotation has previously been detected in
some RS CVn stars (cf. Vogt et al. \cite{hr1099}). These detections
were made by comparing Doppler images having a time difference of up to
several months. This approach is, however, only possible for stars with a
very small differential rotation, since the largest sensible phase 
difference is 0.5. Secondly, it is hard to confirm that the same
particular spots are really present in different images taken months apart.

Strassmeier et al. (\cite{klasu2}) cross-correlated constant latitude slices 
of images to determine the differential rotation of \object{HD~199178}. They 
could not see any effects of differential rotation but stated that this 
``non-detection does not necessarily mean there is no differential surface 
rotation''. We stress that the cross-correlation will provide reliable results
only if there are long-lived spots on different latitudes. That is not the 
case in our images of \object{HD~199178}. Furthermore, the result may be 
biased by artifacts in the images.

Differential rotation can also be detected by combining Doppler images
and time series analysis of photometry. If there is a clear
correlation between the latitude of the main spot and the photometric
period, one could determine the law of the differential rotation. But this 
kind of analysis would require more data than was available for the present 
study. 

\begin{acknowledgements}

The work of T.H. was supported by a grant from the Jenny and Antti
Wihuri foundation, and by Helsinki University research funding for
the project ``Time series analysis in astronomy'' (No. 974/62/98).
I. Ilyin reduced the July 1994 spectroscopic data, participated in the 
observations and gave useful comments on the manuscript. The molecular line 
data were kindly supplied by S. Berdyugina. We also thank the anonymous 
referee, whose suggestions helped to improve the paper. The surface imaging 
temperature maps were calculated using the Cray C94/128 supercomputer at the 
Centre for Scientific Computing, Espoo, Finland. This research has made use 
of the Simbad-database operated at CDS, Strasbourg, France.

\end{acknowledgements}

\end{document}